# SYSTEMS OF OSCILLATORS DESIGNED FOR A SPECIFIC CONSCIOUS PERCEPT

Pavel Kraikivski

*Department of Biological Sciences, Virginia Polytechnic Institute and State University*

*Blacksburg VA, 24061 USA,*

*kraikivski@gmail.com*

As put forward by neuroscientists, the mechanisms of consciousness can be elucidated by revealing correlations between neural dynamics and specific conscious percepts. Recently, I have elaborated on the mathematical formulation for a system of processes that are mutually connected to be isomorphic to a conscious percept of a point in space. Importantly, in such a system, any process can be derived through all other processes that form its complement, or "interpretation." To generate such a solution, I am proposing a dynamical system of oscillators coupled in a manner to preserve the properties of a percept. Specifically, I crafted a dynamical system that retains the mutual relationships among processes, forming an operational map isomorphic to a distance matrix that mimics a percept of space-like properties. The study and results pave a novel way to analyze the dynamics of neural-like (oscillatory) processes with a purpose of extracting the information relevant to specific conscious percepts, which will facilitate the search for neural correlates of consciousness.

## 1. Introduction

Neural correlates of consciousness are the imperative tools to uncover the fundamentals of consciousness. The ultimate goal in the neuroscience of consciousness is to succeed in the search for a minimal set of neuronal events necessary and sufficient for a given state of consciousness[1-9]. Also, various models and approaches have been used to derive the theoretical basis for consciousness, utilizing information and graph theories[6, 10], quantum mechanics[11, 12], logic[13, 14] and operational architectonics[8, 15]. Eventually, if research succeeds in elucidating the fundamental mechanisms of consciousness, engineers will bring artificial computing systems capable of consciousness to life.

It has been summarized in Refs. 15, 16 that the phenomenal patterns are intrinsically self-representing and thus do not need to be presented to anyone. Therefore, the systems capable of consciousness must have a "completeness" property meaning that the processes that form the system must be sufficient to generate interpretation of the pattern by themselves. This property is often used as the common critique addressed to current computing systems when they are compared with the biological neural systems, that computers are not capable of self-interpretation of processes and always require a user to interpret information[16, 17].

To address this issue in this work, I impose the condition that in the modeled system each process can be interpreted and derived through mutual relationships with other processes that form its complement. For example, we cannot think about a point without a space, and the space is a collection of all other places where we think the point can exist. Thus, each point is interpreted through particular relationships with all other points in space. We can think about all possible places for a point due to our perceptual experience of objects in different locations. This experience is recorded in our memory which reflects the relationships among possible locations. My assumption is that a stimulus gets an interpretation in the brain through mapping of a corresponding internal process into a set of processes that correspond to all other possible locations of stimulus perceived and recorded in the brain before. Such set of processes form an internal operational space isomorphic to the stimuli space. Thus, each process is not alone but with the set of other processes that bound to it. And as we get more experience, the more processes are involved in such interpretation.

Thus, generalizing the idea, if a dynamical system produces the set of processes $\vec{P} = (p_1, p_2, \ldots, p_n)$ in response to a set of external inputs (stimuli), then any process $p_i$ (or a subset of processes) that corresponds to a single input must have an interpretation through relationships between the process $p_i$ and all other processes in the system[18]. The process $p_i$ communicates with other processes by recruiting the memory of the system which reflects the mutual relationships between process $p_i$ and all other processes. Such memory must be gained through the training of the system and reflects the spatial and time order between different stimuli that the system can perceive during the training. Thus, a relationship map for internal processes in the system must be isomorphic to mutual relationships among experienced percepts. The idea of isomorphism between neural activity and phenomenal space has been proposed to be central for consciousness realization[7, 8, 15, 19, 20]. Also, for a system to synchronize many processes into a dynamical pattern isomorphic to a complex conscious percept with different cognitive features and phenomenal aspects, it might be essential for the system to be capable of achieving temporo-spatial alignment and globalization among processes as proposed in the temporo-spatial theory of consciousness [21, 22].

The relationships between a process and all other processes can be represented by a matrix $A$[18]. Each element of the matrix represents a connection between a process $p_i$ and a process $p_j$. This connection reflects the relationship among corresponding external stimuli (e.g. spatial or time separation of stimuli) that the system perceives to generate these processes. Also, the dynamical system must preserve the map of mutual relationships between processes $A$ in a steady-state manner. For examples, the processes could be oscillating but the relationships between amplitudes (or other features) of processes must be fixed to preserve $A$. This property is essential to bring the theory in agreement with our experience that a conscious percept can persist unchanged in time as long as the corresponding neural activity remains stable.

The assumption that interpretation in a system of processes arises from mutual relationships among processes is in line with the phenomenological observation that "something is determined as opposed to an other" as elaborated on in Hegel's *Science of Logic*[14] and is also in agreement with the law of the unity and conflict of opposites formulated in dialectical materialism[23]. I assume that we always interpret things through their relationships with other things and that the same relationships are preserved on the level of processes as among stimuli that cause these processes. For example, the color black or darkness has meaning for us because we also know what it is like to experience the presence of light, the color white and all possible gray levels that form a space (a set of opposites) in which the absence of light is interpreted. And if we would never experience all possible gray levels, then our experience of darkness would be different from if we have such experience.

One of the popular examples demonstrating the importance of the actual experience for color perception is a hypothetical experiment with a person named Mary living isolated in a colorless world but studying all physical properties of a color[24, 25]. Even knowing all physical properties of color, Mary fails to experience the color unless the actual color is shown to her. However, I further assume that even if the color red will be shown to Mary she will not experience redness of red the same way as we do. Mary will perhaps experience red as a kind of grey unless she learns also about all possible colors, and her memory will store all relationships among different colors. The indirect support of this hypothesis has been demonstrated in physiological experiment described in Ref. 26 with new born kittens raised in the experimental environment with either horizontal or vertical black-and-white stripes for five months. Then, when kittens are placed in a normal environment, they fail to detect objects or contours that were aligned in the opposite way to their previous environment[26].

For the mathematical framework described in this work, this means that the system of processes and connectivity $A$ change during learning (presents of new stimuli). And thus the interpretation of each process $p_j$ will also alternate during learning unless no new input can make a qualitative difference (change symmetry or structure of matrix $A$) for relationships between $p_j$ and other processes. However, I do not propose any system that can be trained or any training method for a system of processes. I assume a fixed set of processes and wire them to interpret a specific percept (like seeing a point-like object or a thing) such that the matrix $A$ will reflect the relationship between the object (thing) with all other things that we can characterize as opposite or different by some property. For cases when the analytical solution for matrix $A$ cannot be derived, simulation results obtained using Dehaene–Changeux model[27, 28] trained for complex percepts can provide data to derive the relationships among processes in neural-like dynamical systems. The experiments that can validate the "completeness property" of a system in which any process is derived through other processes forming its complement and the data needed for derivation of matrix $A$ will be discussed in this manuscript.

Formulation based on mutual information used in Integrated Information Theory (IIT)[10, 29] shares similarities with my interpretation of states through the mutual relationships among them. Furthermore, as explained in IIT[10], the occurrence of a given conscious state implies an extremely rapid selection among a repertoire of possible conscious states or when a system reduces uncertainty among a repertoire of possibilities in a probabilistic way. In my deterministic approach, all states must exist (all connected processes related to different stimuli learned by a system are running) at a time when a conscious percept is realized. And it is a dynamical interplay between a state (associated with current stimulus) and other states (in memory due to past inputs) that is involved in realization of the conscious percept. The properties encoded in mutual relationships among processes become or emerge as a conscious percept. These properties are developing when a dynamical system of processes continues to exhibit them in a steady-state manner, and thus the corresponding conscious percept continues to be present in the system.

In this manuscript, I design a dynamical system with an internal functional organization that is isomorphic to a specific conscious percept. I present the solution for the system and show how the information on internal functional organization is encoded in the amplitude pattern of oscillating processes. The goal is to present a simple model that can teach us how to recover the functional organization of the system from the amplitude patterns of neural firings obtained from the high-resolution imaging techniques used in the experimental search of the neural correlates of consciousness.

## 2. Model: wiring the mechanism for a specific percept

I aimed to derive a dynamical system that retains the mutual relationships among processes such that any process can be obtained through the others. Thus, for a set of processes $\vec{P} = (p_1, p_2, \ldots, p_n)$, the dynamical system must retain the relationship $\vec{P} = A\vec{P}$, where $A$ is the hollow matrix, with the diagonal elements all equal to zero. The structure of $A$ is time-independent and represents the memory of the system, and the elements of the matrix are the couplings between processes and could mimic the strengths of the synaptic connections in a neural system. Also, I will use the term the internal functional organization of processes referring to the structure of matrix $A$.

As an example, I will use the Euclidean distance matrix to mimic space-like characteristics for the system of the process:

$$p_i = \sum_{j=1}^{n} (i-j)^2 \varepsilon \, p_j \tag{2.1}$$

and thus $A = \begin{pmatrix} 0 & \varepsilon & \cdots & (n-1)^2 \varepsilon \\ \varepsilon & 0 & \cdots & (n-2)^2 \varepsilon \\ \vdots & \vdots & \ddots & \vdots \\ (n-1)^2 \varepsilon & (n-2)^2 \varepsilon & \cdots & 0 \end{pmatrix}$, $\tag{2.2}$

where $\varepsilon$ is a constant that will be determined. Mathematically $\varepsilon$ can be interpreted as eigenvalue if the scalar $\varepsilon$ is pulled out of the matrix, but it also plays the role of a scaling parameter for the "distance" measure between processes in (2.1) formulation. Here, the assumption is that if a neural-like system is trained using spatial patterns, e.g. a point in space at different positions $(i, j)$ separated by the distance $x_{ij} = |x_i – x_j|$, where $x$ is a coordinate of a point like stimulus along the line, then the positions and distances will be encoded in the neural processes such that their internal organization will be isomorphic to Euclidean space. Thus, the relationship between processes of the neural-like system can be represented by the matrix $A$ which is isomorphic to Euclidean distance matrix. The graphical representation of coupled processes is shown in Fig 1.

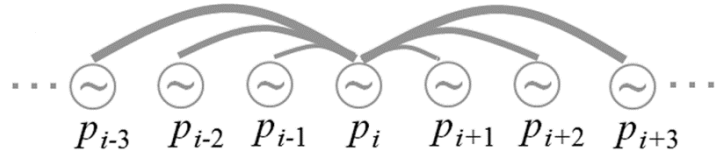

Fig 1. The graphical illustration of mutual relationships among executed processes for the symmetric case corresponding to equation (2.1). The gray lines that link processes (circles with a wave inside) represent the mutual interactions of process $p_i$ with others. The thickness of these lines represents the variation of couplings between processes.

I aim to design a dynamical system that is oscillatory and preserves the mutual relationships among processes, $A$. The structure of matrix $A$ must be an intrinsic property of the dynamical system. A set of ordinary differential equations satisfying such conditions can be written in the following form:

$$\frac{d\vec{P}}{dt} = A\vec{P} - (\vec{Z} + \vec{P})$$

$$\frac{d\vec{Z}}{dt} = \vec{P} \quad (2.3)$$

The system (2.3) has a solution $\ddot{\vec{P}} = A\vec{P}$ with oscillating $\vec{P} = \vec{Q}\cos(\omega t) + \vec{R}\sin(\omega t)$, where $\vec{Q} = (Q_1, Q_2, ..., Q_n)$ and $\vec{R} = (R_1, R_2, ..., R_n)$ are amplitudes for processes $(p_1, p_2, ..., p_n)$ to be defined by initial conditions. Thus, system (2.3) preserves the mutual relationships between processes (2.1). The purpose of Z is to create oscillating carriers of processes P. The dynamical system can be of any kind; it must only preserve the functional organization among processes: $\ddot{\vec{P}} = A\vec{P}$. This assumption is along with David Chalmers's principle of organizational invariance according to which if a system has conscious experiences, then any other system with the same fine-grained functional organization will have qualitatively identical experiences[30]. In my work, the internal organization among processes is represented by the matrix $A$, and it is the functional organization if **A** is isomorphic to a specific conscious percept and the structure is maintained in the dynamical system such that $\vec{P} = A\vec{P}$ in a steady-state.

Equations (2.3) can be also rewritten in terms of Z variables:

$$\frac{d^2\vec{Z}}{dt^2} + \frac{d\vec{Z}}{dt} + \vec{Z} - A\frac{d\vec{Z}}{dt} = 0 \quad (2.4)$$

Equations (2.4) describe damped harmonic oscillators mutually coupled through rates of change in their dynamics.

The underlying oscillatory motifs in the system (2.3) are very common for biological systems[31]. Fig. 2A shows a network motif that is a basic unit of wiring representing the model system (2.3). With a positive feedback of $p_i$ on itself the motif in Fig. 2A would represent an oscillatory network motif extensively studied in systems biology[31]. With a negative feedback ($p_i$ on itself) oscillations are damped. The negative feedback, however, is vanishing when $\vec{P} = A\vec{P}$ and oscillations are then maintained due to supporting inputs from other processes shown in Fig. 2B. The wiring complexity demonstrated in Fig. 2B is similar to what is used in the Dehaene–Changeux model[27, 28], yet it is just designed to model a simple percept.

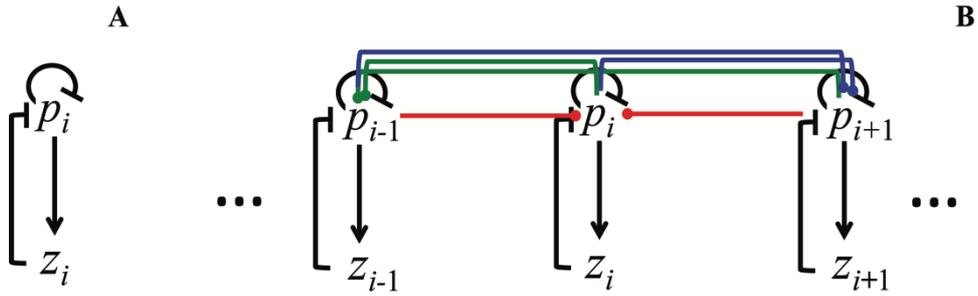

**Fig 2. Wiring diagrams:** (A) shows network a motif which would represent the system (2.3) if processes $p_i$ would be disconnected (parameter ε=0), where arrow-headed lines represent activating interaction (positive influence) and bar-headed lines represent inhibiting interaction (negative influence). (B) shows the wiring between neighboring processes in (2.3), where the dot-headed line can represent positive or negative influence depending on the sign of ε parameter and whether processes oscillate in phase or oscillate with the half period phase shift (constrictive or distractive interference). Different line colors are used for tracking purpose. Red lines represent interactions between $p_{i-1}$, $p_{i+1}$ and $p_i$, green lines represent interactions between $p_i$, $p_{i+1}$ and $p_{i-1}$, blue lines wire $p_{i-1}$, $p_i$ with $p_{i+1}$.

## 3. Results

In general, for arbitrary **A** the system (2.4) represents non-linear system and oscillators are coupled. However, if processes are converging to the solution $\vec{P} = A\vec{P}$ then (2.4) represents uncoupled harmonic oscillators. In such special case, the damping term $\frac{d\vec{Z}}{dt}$ is compensated by the positive feed determined by the change in oscillations of all other oscillators $A\frac{d\vec{Z}}{dt}$. Thus, the oscillators can maintain the property of the system: $\vec{P} = A\vec{P}$ that each process is described through other processes in a specific way that is defined by the matrix **A**. In this case, processes will be eigenvectors of matrix **A**. However, if processes are perturbed such that $\vec{P} \neq A\vec{P}$ then oscillators are become coupled again and the equation (2.4) will described coupled oscillators. In the case of $\vec{P} \neq A\vec{P}$, the system (2.4) will be either in a damping mode with oscillations that have decreasing amplitude in time or an amplifying mode with oscillations growing in amplitude. I will assume that the system is perfectly tuned to $\vec{P} = A\vec{P}$. In a real neural system the ability of maintaining the solution $\vec{P} = A\vec{P}$ can be achieved with additional negative and positive feedbacks from other processes.

In this manuscript, I don't aim to claim that the system (2.4) is describing a real neural system and also I don't aim to explore all mathematical properties of the system (2.4). Instead, I hypothesize that if a system is capable to maintain specific mutual relationships among processes that are functionally organized isomorphic to a conscious percept then such a system is capable of experiencing this percept as long as it maintains such relationships among processes. And my second phenomenological hypothesis is that the consciousness happens when the dynamical system brings the specific relationships among processes into existence meaning that the relationship $\vec{P} = A\vec{P}$ must happen in the dynamical system and it must be maintained during the period of conscious perception. Thus, I aim to provide an example and demonstrate that it is possible to build a dynamical system with emerging specific organization among processes $\vec{P} = A\vec{P}$.

I will also show that if the system is tuned to maintain the relationships between processes such that $\vec{P} = A\vec{P}$ then it is stable to perturbations in the initial conditions. The initial conditions can be considered as different inputs to the system and the matrix $A$ as an internal structure (wiring among processes) of the system. The initial input value for a process mimics the excitation of a process due to a stimulus. Perhaps, for the real neural system, even a simple stimulus like a point in space will initiate the set of action potentials however this set can be represented as a single process $p_i$ that is used in the formulation (2.1). When the position of a stimulus changes then the activity of another group of neurons in neural system will be represented as a single processes $p_j$ which will be coupled with $p_i$ through Hebbian-like learning mechanism according to which the synaptic connections between previous group of neurons and the new group of neurons would strengthen[32]. Thus for the neural system trained with spatial patterns, Hebbian learning must result in the connectivity map among firing neurons isomorphic to the matrix $A$.

### 3.1. *The relationships among processes are encoded in amplitude patterns*

I consider several specific examples of the systems containing different numbers of processes and different initial conditions to demonstrate how the relationship $\vec{P} = A\vec{P}$ is maintained in the dynamical system and how it is encoded in the dynamical pattern of processes.

For the system of two processes $\vec{P} = (p_1, p_2)$, the variables $(z_1, z_2)$ are described by the following two equations:

$$\frac{d^2 z_1}{dt^2} + \frac{dz_1}{dt} + z_1 - \varepsilon \frac{dz_2}{dt} = 0$$

$$\frac{d^2 z_2}{dt^2} + \frac{dz_2}{dt} + z_2 - \varepsilon \frac{dz_1}{dt} = 0$$

and $(p_1, p_2)$ can be obtained by taking the time derivatives of $(z_1, z_2)$ as defined by the second equation in (2.3). The solutions of this system of two oscillators in the form of $(p_1, p_2)$ are shown in Fig 3A. For ε = +1, the processes $(p_1, p_2)$ oscillate in phase and for ε = −1, the processes $(p_1, p_2)$ oscillate with the half period phase shift as shown in Fig. 2A. If the magnitude of the parameter ε is even just slightly bigger than one then the amplitude of the oscillations will continue to grow each oscillation cycle. In this case, the feeding term $\varepsilon \frac{dz_i}{dt}$ in the equations is dominating over the damping term $\frac{dz_1}{dt}$. If the magnitude of ε is less than one then oscillations will be damped.

The system of *n* processes is modeled by *n* coupled oscillators described by *n* ordinary differential equations, which are written in the vector form (2.4).

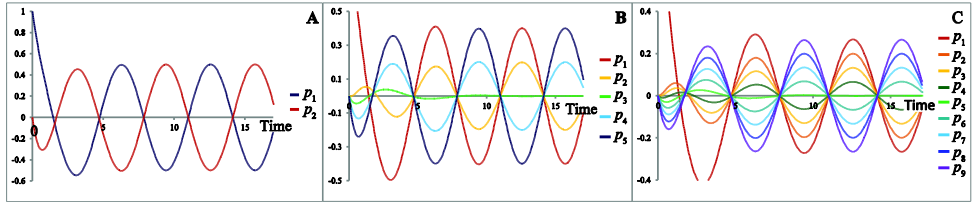

**Fig 3**. **The solutions of (2.3) for the systems consisting of two (A), five (B) and nine (C) processes**. All processes start at zero, except the first process, which has an excitation value of 1 at zero time. The Figure shows the relative organization of processes in time, which reflects the mutual relationships among processes $\vec{P} = A\vec{P}$. The values of ε used to generate solutions are $\varepsilon = -1$ for (A), $\varepsilon = -1/20$ for (B) and $\varepsilon = -1/120$ for (C).

The system (2.4) is written in the form of oscillator equation that has a simple interpretation. However, the system (2.3) contains only first order ordinary differential equations (ODEs) that can be easily managed numerically using various open-source software platforms designed to solve first order ODEs. All graphical solutions of ODEs in this manuscript were obtained using the parameter estimation toolkit (PET) developed in John Tyson's computational cell biology lab (which is available at http://mpf.biol.vt.edu/pet/). The system (2.3) can be solved for any number of processes; see the solution examples in Figs 2-4. Also, the parameter ε in $\vec{P} = A\vec{P}$ has several values to satisfy the equation $\det(A - I) = 0$, where det is the standard notation for the determinant of the matrix. Figs 2A-2C, 3A and 4A show solutions for the systems with two, five, nine, ten and four processes, correspondingly. The solutions in Figs 2 and 3 are obtained for the initial conditions when only one process has a non-zero initial value: $p_1(t = 0) = 1$, and all other $p_i(t = 0) = 0$ for $i \neq 1$, to mimic the input of a single dot in space. The solutions of equations (2.3) for different initial conditions (inputs) for the system consisting four processes are shown in Fig 5.

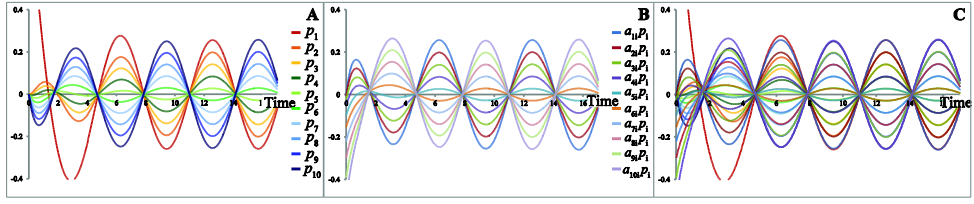

**Fig 4. The dynamics of the system of ten processes.** (A) The evolution of processes $p_j$ that satisfy equation (2.3). (B) The time dependence for the sum of processes that matches with process $p_j$ as shown in (C). To simplify the summation notation, the sum is assumed over the range of repeating the matrix index $a_{ji}p_i = \sum_{i=1}^{10} a_{ji}p_i$, where $a_{ji}$ are the elements of the hollow matrix $A$. Graphs in (A) and (B) are plotted together in (C) to show the dynamics of how process $p_j$ matches with the weighted sum of other system processes $\sum_{i=1}^{n} a_{ji}p_i$, through which it thus can be defined. The parameter $\varepsilon = -1/165$ was used to generate these solutions.

The results shown in Fig 3 indicate that the information on mutual couplings between processes is encoded in amplitude patterns. The mutual couplings among processes are isomorphic to a conscious percept of space-like characteristics (points and distances). Thus, the model provides the algorithm of how to correlate the amplitude patterns generated by a system of processes (one that mimics neural systems) with a specific type of conscious percepts. Importantly, the dynamical system (2.3) or, equivalently, (2.4), has the property that any internal process is also defined by its complement, which is demonstrated in Fig 4.

The time plots in Fig 4B show the dynamics for $A\vec{P}$, the processes that propagate through a "memory lens," $A$. Fig 4C demonstrates the matching dynamics between processes and their complements. This proves that in this dynamical system, any process $p_j$ can be defined through its complement $\sum_{i=1}^{n} a_{ji}p_i$, where $a_{ji}$ are the elements of the hollow matrix $A$. As shown in Fig 4C for the system of ten processes, $p_j = \sum_{i=1}^{10} a_{ji}p_i$ at steady state. For example, a particular process $p_1$ is defined through processes $p_2,\dots,p_{10}$ as $p_1 = \sum_{i=1}^{10} a_{1i}p_i = a_{12}p_2 + a_{13}p_3 + \cdots + a_{110}p_{10}$.

### 3.2. *Convergence of the internal and external worlds*

The initial conditions can be interpreted as inputs set by the presence of an external stimulus to the system with internal structure defined by $A$. As a first response to the external stimulus, the dynamics of internal processes are affected by the matrix $A$, and $\vec{P} \neq A\vec{P}$, see dynamics of $\vec{P}$ and $A\vec{P}$ in Fig. 4 at small times. At these times, when $\vec{P} \neq A\vec{P}$, the system (2.3) behaves as non-linear system of coupled oscillators. After some time, the processes converge such that $\vec{P} = A\vec{P}$, hence each process obtains the interpretation from the system. The system must already have internal structure represented by matrix $A$ that the neural-like system could get from training or experience.

Importantly, there is only a trivial solution $\overrightarrow{P(t)} = \vec{0}$ for the initial condition $p_i(t=0) = 0$ for all $i = 1,\dots,n$. The behavior of a system of four processes is investigated for different initial inputs and is shown in Fig 5. The dynamics of how a process in the system matches with the intrinsic dynamics of other processes $A\vec{P}$, as well as amplitude patterns, depend on the initiation inputs for the other processes. The results reflect how the dynamical system interprets the input pattern with a given internal functional organization of processes, *A*.

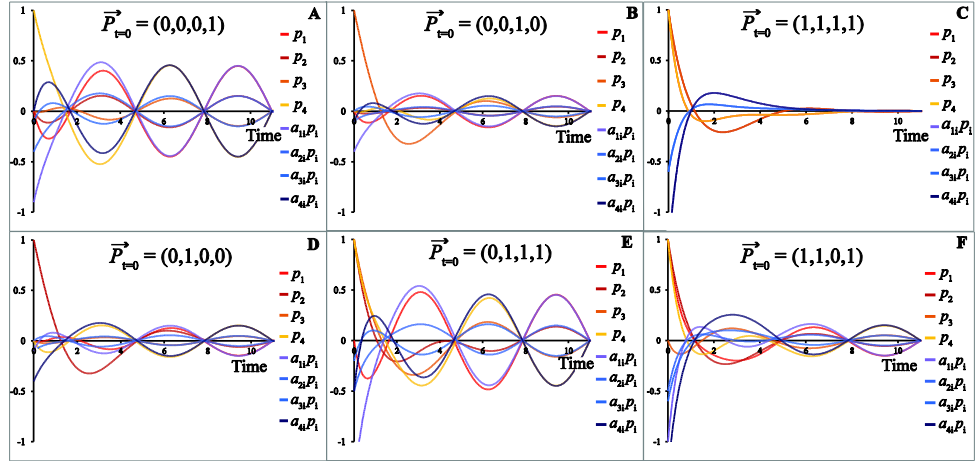

**Fig 5. The change in amplitude patterns depending on initial conditions**. Figs (A) ‐ (F) correspond to different initial values of processes in the vector $\vec{P}$, which is specified in the header of each Figure at zero time. The solutions for a system consisting of four processes are shown, as well as the dynamics for $a_{ji}p_i = \sum_{i=1}^{4} a_{ji}p_i$, to show how the processes $p_j$ match with thier corresponding complements, the index $j = 1,2,3,4$. The parameter $\varepsilon = -0.1$ was used to generate solutions.

The set of $\varepsilon$ parameters (-1, -0.25, -0.1,-0.05,-1/35,-1/56,-1/84,-1/120,-1/165) for the system, consisting from two to ten processes, were used. However, there are also other solutions (eigenvectors) that satisfy $\vec{P} = A\vec{P}$ with different $\varepsilon$ scaling factors, which can have also positive values.

The model provides a good example of how a system's internal functional organization of processes (the specific dynamical relationship among processes, $\vec{P} = A\vec{P}$) is encoded in the amplitude patterns of processes that can be measured by such high-resolution imaging techniques as functional magnetic resonance imaging and computed tomography. The study of such deterministic model systems reveals the mathematical algorithm that could help to extract the relevant information on specific conscious percepts from neural patterns and, thus, can be instrumental in the search for neural correlates of consciousness. Importantly, some system properties (properties of matrix *A*) such as symmetry and the organization of mutual relationships among processes are invariant to reduction in the number of a system's processes (e.g. the distance matrix *A*, the symmetry $a_{ji} = a_{ij}$, and it remains isomorphic to Euclidean space for different number of processes *n* in (2.1)) This idea mimics the brain property that although some neural cells are lost, the conscious percept of space remains the same.

## 4. Experimental Testing

The model results predict that information perceived by the neural system is encoded in amplitude patterns. This is consistent with an observation made using electroencephalograms (EEG) of the olfactory bulb of rabbits and cats. It has been demonstrated that information concerning the identity of a particular odor was not carried by the temporal shape of any particular EEG wave but by the spatial pattern of the EEG amplitude across the entire surface of the olfactory bulb[33, 34]. However, according to the model, the temporal synchronization among processes is required such that all mutual relationships of amplitudes among processes are preserved in steady-state manner. The model is in agreement with Libet and coworkers' earlier studies demonstrating that direct microelectrode stimulation of the somatosensory cortex must exceed 200 ms in order for the subjects to perceive that percept[35, 36]. The suggestion that consciousness of input is preceded by a period of preconsciousness processing is also supported by Neely's works in Refs. 37, 38 , and a common estimate of preconscious processing time is in the order of 250 ms.

The model results suggest a tool to analyze the data that can help experimental neuroscientists identify neural correlates of consciousness. Data obtained using EEG, high-resolution imaging techniques such as functional magnetic resonance imaging (fMRI) and computed tomography on responses to a sequence of position stimuli could be mapped in such a way that any response to a particular stimulus could be related to all other responses (to other position stimuli). Then the matrix (2.2) can be derived from such experimental data. The map of neuronal correlates generated in the human brain in response to such a simple stimulus as the position of a point-like object along a line would be one of the first tests that can directly support the mathematical model presented in this manuscript. The assumption that the neural system must have the completeness property seems to be in logical agreement with the mechanisms of plasticity and long-term potentiation observed in the human nervous system[39]. The long-term potentiation reflects a persistent strengthening of connections between two neurons based on their recent patterns of activity. The training of the neural system with a temporal sequence of position stimuli would most likely result in a neural connectivity map that mimics stimuli patterns and will be isomorphic to the distance matrix for the spatial patterns. In this case, each neuronal process will be defined through mutual relationships with other processes generated in response to spatial stimuli. This can be directly proven using the fMRI technique that measures the correlation between the fMRI response and a stimulus[40], and thus could reconstruct the dynamical map generated in response to a sequence of spatial stimuli.

The aim to define a model system dedicated to a particular percept aligns with the experimental search for a minimal neural mechanism sufficient for any one specific conscious percept[4]. The mathematical formulation of a system designed for a particular conscious percept was motivated by the data from fMRI technique, the data on patients with abnormal damage of brain tissue (lesions) and by the fact that some living species with small neural systems can generate very comprehensive neural processing to various environmental cues. fMRI data show that brain activity is localized to a specific area in response to a simple stimulus[40]. The data on patients with brain lesions also support that neural systems smaller than the whole brain are capable of consciousness; for example, split-brain operations yield two separate subjects of conscious experience[41]. Relatively small neural systems exist in some living multicellular species. For example, C. elegans has two distinct and independent nervous systems: a large somatic nervous system (282 neurons) with about 6400 synaptic connections and a small pharyngeal nervous system (20 neurons)[42]. Yet, such relatively simple and compact nervous systems are capable of processing environmental cues and initiating such complex responses as moving toward or away from chemicals, odorants, temperatures, and food sources[43]. The test on completeness of a neural system could be used to classify the operational complexity of neural systems in various living species.

## 5. Discussion

Overall, I have used a bottom-up approach to design a simple dynamical system for a specific conscious percept. The results can pave a novel way to analyze data on neural oscillations. Mapping neural processes into a matrix that connects each process to other processes, $\vec{P} = \boldsymbol{A}\vec{P}$, will reveal essential properties of conscious percepts, which will be more evident in the matrix form (e.g. symmetries). In addition, such mapping will be useful to identify a complete or minimal set of processes that corresponds to a specific conscious percept.

The simulation results of a dynamical system of processes indicate that the information on mutual couplings between processes that are chosen isomorphic to a conscious percept of space-like characteristics (points and distances) is encoded in amplitude patterns. Indirectly, this is consistent with an observation that electroencephalograms (EEG) of the olfactory bulb of rabbits and cats demonstrated that information concerning the identity of a particular odor was not carried by the temporal shape of any particular EEG wave but by the spatial pattern of the EEG amplitude across the entire surface of the olfactory bulb[33, 34]. The spatial information perceived by the neural system could also be encoded in amplitude patterns.

In this work, I have aimed to design a dynamical system that preserves a particular operational map of mutually connected processes. The map provides an interpretation to any process in the system through the other processes that form its complement. Also, the interpretation is different from higher-order thought idea elaborated by David Rosenthal in Ref. 44 because all processes are involved and not separable (all are needed for completeness in the system) and the state is perceived in the properties that dynamical system is exhibiting. The process evokes other processes bound to it and other processes evoke the one realizing oscillatory interplay between one and others.

In this work, I have crafted a dynamical system that retains the mutual relationships among processes such that the resulting map will be equivalent to the distance matrix that mimics a percept of space-like properties. This example was used to demonstrate the idea of completeness in the dynamical system. Using a similar design strategy, other dynamical systems can be crafted to retain specific properties of the operational space consisting of mutually connected processes. The asymmetric map of mutually connected processes, which can be interpreted as a direction in space, has been elaborated in Ref. 18. It should also be possible to build a dynamical system retaining the operational map that is isomorphic to color-space, or space corresponding to odor or sound percepts. The philosophical aspects of consciousness in relation to such systems are also discussed in Ref. 18.

The deterministic model with an exact solution can be used to derive the algorithm to solve the reverse problem of extracting couplings between processes from the amplitude patterns. Representing couplings in a matrix form will offer a new way to analyze encoded information. Thus, the algorithm will provide a novel tool to analyze neural firing patterns from such high-resolution imaging techniques as functional magnetic resonance imaging and computed tomography.